\documentclass[twocolumn]{aastex631}

\usepackage{derivative}
\usepackage{amsmath}
\usepackage{ulem}

\shorttitle{Formation of inner planets}
\shortauthors{Guo \& Kokubo}
\graphicspath{{./}{figures/}}
\accepted{August 21, 2023}

\begin{document}

\title{Formation of inner planets in the presence of a Cold Jupiter: orbital evolution and relative velocities of planetesimals}

\correspondingauthor{Kangrou Guo}
\email{carol.guo@sjtu.edu.cn}

\author[0000-0001-6870-3114]{Kangrou Guo}
\affiliation{Tsung-Dao Lee Institute, Shanghai Jiao Tong University, 520 Shengrong Road \\
Shanghai 201210, People's Republic of China}

\affiliation{The University of Tokyo, 113-8654, 7 Chome-3-1 Hongo \\
Bunkyo City, Tokyo, Japan}


\author[0000-0002-5486-7828]{Eiichiro Kokubo}
\affiliation{The University of Tokyo, 113-8654, 7 Chome-3-1 Hongo \\
Bunkyo City, Tokyo, Japan}

\affiliation{National Observatory of Japan, 181-8588, 2 Chome-21-1 Osawa \\
Mitaka City, Tokyo, Japan}

\begin{abstract}

We investigate the orbital evolution of planetesimals in the inner disk in the presence of nebula gas and a (proto-) cold Jupiter.
By varying the mass, eccentricity, and semi-major axis of the planet, we study the dependence of the relative velocities of the planetesimals on these parameters. 
For classic small planetesimals ($10^{16}-10^{20}~$g) whose mutual gravitational interaction is negligible, gas drag introduces a size-dependent alignment of orbits and keeps the relative velocity low for similar-size bodies, while preventing orbital alignment for different-size planetesimals.
Regardless of the location and the mass ratio of the planetesimals, increasing the mass and eccentricity or decreasing the orbital distance of the planet always leads to higher relative velocities of planetesimals. 
However, for massive planetesimals, the interplay of viscous stirring, gas damping, and secular perturbation results in lower velocity dispersion of equal-size planetesimals when the planet is more massive or when it is located on a closer or more eccentric orbit.
The random velocities of such planetesimals remain almost unperturbed when the planet is located beyond Jupiter's current orbit, or when it is less massive or less eccentric than Jupiter. 
Unlike small planetesimals, such large planetesimals can grow in a runaway fashion as in the unperturbed case.
Our results imply that the presence of a cold Jupiter does not impede the formation of inner rocky planets through planetesimal accretion, provided that the planetesimals are initially large.


\end{abstract}



\section{Introduction} \label{sec:intro}

The formation of rocky or terrestrial planets like our Earth is generally acknowledged to begin from dust aggregation into kilometer-sized planetesimals, which are the building blocks of planetary bodies. 
Planetesimals then gravitationally accumulate in the "runaway growth" mode \citep[e.g.][]{WETHERILL1989,Kokubo_Ida_1996} followed by "Oligarchic growth" mode \citep{Kokubo_Ida_1998} to form Lunar- to Mars-sized planet embryos.
The last stage is featured by giant impacts among planet embryos and eventually produces full-sized rocky planets.

The stage of planetesimal accretion is very sensitive to external gravitational perturbations, because they can excite the random velocity of a swarm of planetesimals and reduce the growth rate.
A stellar companion, for example, can be a source of such perturbations. 
A great number of studies have investigated the influence of stellar binarity on the formation of planets in binary star systems and explored the conditions (e.g. binary parameters, orbital distance) under which planet formation is possible \citep[e.g.,][]{Marzari_Scholl_2000, Scholl_et_al_2007, Marzari_et_al_2008}.
Apart from stellar companions, giant planets can also be a source of perturbation and affect the formation process of rocky planets.
Collisional evolution of planetesimals in a gas-free disk perturbed by a giant planet embryo has been investigated in previous studies \citep[e.g.,][]{Thebault_1998, Charnoz_et_al_2001, Charnoz2001}.
More complicated scenarios with the gas disk in presence have been investigated by other studies\citep[e.g.,][]{KW00, Kortenkamp_et_al_2001, Thebault_2002}.
\citet{KW00} showed that under the coupled effect of gas drag and the perturbations from Jupiter and Saturn, the encounter velocities of similar-sized planetesimals near 1 au is significantly reduced by orbital alignment so that the growth of terrestrial planets can be facilitated.
Such an effect of velocity dip for similar-size planetesimals is identified as "Type II" runaway growth, in which the effects of dynamical friction in the classical "Type I" runaway growth are mimicked by the size-dependent phasing of orbital elements \citep{Kortenkamp_et_al_2001}.

Despite the generality of their results, the initial conditions in KW00 and \citep{Kortenkamp_et_al_2001} that assume fully formed Jupiter and Saturn (possibly by disk instabilities) at their current location with their current orbital elements before the accumulation of inner planetesimals seem open to discussion.
Although the \textit{disk instability} model \citep[e.g.,][]{Kuiper_1951,Boss_2000} provides a good route to form gas giants on a short timescale, it is less favored for explaining the formation of the gas giants in the Solar System when compared with the \textit{core accretion} \citep[e.g.,][]{Mizuno_1980,Pollack_et_al_1996} model.
Rather, the \textit{disk instability} model is preferable for producing massive companions (e.g., brown dwarfs and cold giant planets) on wide orbits, which are usually detected by direct imaging \citep[e.g.,][]{BOLEY2010509,Boss_2011,Fletcher_2019}.
If Jupiter and Saturn had formed through core accretion, their masses would have been smaller during the early formation of terrestrial planets, and their eccentricities could also deviate from the present values.
Furthermore, the locations of these gas giants could have shifted due to orbital migration.
To account for possible inward migration, KW00 and \citet{Kortenkamp_et_al_2001} also included the case when Jupiter is located at 6.2 au and Saturn at 10.5 au, in addition to the case where these two planets are at their current locations.
However, as suggested by the Grand Tack model, these gas giants could have undergone inward migration and then outward \citep{Walsh2011}.
Therefore, it is also necessary to investigate the case where Jupiter is on a closer orbit than its current location.

By varying the mass, eccentricity, and semi-major axis of the outer giant planet, we can generalize the models in previous studies \citep[e.g.,][]{KW00,Kortenkamp_et_al_2001,Thebault_2002} to exoplanetary systems.
The previous studies usually construct the models based on a certain system.
For example, \citet{KW00} and \citet{Kortenkamp_et_al_2001} considered the inner planetesimals perturbed by Jupiter and Saturn; \citet{Thebault_2002} studied the formation of terrestrial planets in 47 UMA and Epsilon Eridani, taking into account two different scenarios where the outer planet was fully formed when the inner planetesimals were in early phase of accumulation and where the giant planet reached its final mass later on when lunar-sized planet embryos had already formed in the inner disk.
This leaves us room to explore the dependence of these results on multiple parameters of the perturbing planet.
Many numerical studies have investigated the dependence of the planetesimal velocities on binary parameters (e.g., binary eccentricity, inclination, and mass ratio) in the circumprimary disk \citep[e.g.][]{THEBAULT2006193, Xie_2009, Silsbee2021}, a model configuration similar to the inner planetesimal disk perturbed by a cold giant planet in a single star system.
However, the range of these parameters and thus the strength of the active forces (e.g., gravitational perturbation, gas drag) are quite different when we consider a planet as the perturber instead of a stellar companion. 
Therefore, the results of the numerical studies on circumprimary planet formation cannot be applied to the model of interest here, and numerical simulations based on a model suitable for the formation of inner planets in a single star system with an early-formed cold giant planet are necessary.
\citet{Rafikov_2014I,Rafikov_2014II} have analytically explored the secular dynamics of planetesimals in circumprimary disks taking into account the gravity of the eccentric disk, perturbations of the stellar companion, and gas drag. 
In order to quantitatively estimate the degree of orbital alignment and the relative velocities of planetesimals during the very early stage of accretion, we perform numerical simulations with suitable parameters to mimic an inner planetesimal disk perturbed by a cold giant planet.

In this paper, we generalize the numerical results of previous studies to account for the case where Jupiter has formed not necessarily through disk instabilities but also from the conventional core accretion scenario, and to apply the results to extrasolar systems by varying the mass, eccentricity, and semi-major axis of the (proto-) cold Jupiter (CJ).
In particular, we focus on the evolution and distribution of the orbital elements and the relative velocity of planetesimals under the coupled effect of gas drag and the perturbation from a CJ in the outer disk.

This paper is arranged as follows:
In Section \ref{sec:method} we introduce our model setup and explain how we calculate the relative velocity of planetesimals. 
In Section \ref{sec:results} we present our results from numerical simulations and show how they depend on the planet parameters.
We then discuss about our results and their implications in Section \ref{sec:discussion}. 
Finally, we summarize this paper in Section \ref{sec:summary}.

\section{Method} \label{sec:method}

We numerically calculate the orbital evolution of planetesimals under the coupled effect of gas drag and perturbations from an external giant planet.
We investigate the dependence of the encounter velocity of planetesimals on the mass, eccentricity, and semi-major axis of the planet by varying these three parameters in our models.
The numerical integrator we use is a fourth-order Hermite integrator \citep{Kokubo_Makino_2004}.
More details of the integrator can be found in Section 2.2 in  \citet{Guo_Kokubo_2021} (hereafter GK21).

\subsection{Model Setup} \label{subsec:models}
This section describes how the ranges of parameters are chosen and how the models are arranged for comparison.

\subsubsection{planet mass}
In order to account for the case where Jupiter has not yet formed while the terrestrial planets start accreting, we need to consider a planet smaller than the current Jupiter.
Planet embryos could have formed from planetesimal rings created by pressure bumps in the protoplanetary disk; in an inner ring where the total mass in planetesimals is $\sim 2.5~M_{\oplus}$, planetesimals grow to roughly Mars-mass planetary objects in $\sim 1-3$ Myr via planetesimal accretion near 1 au \citep{Izidoro_et_al_2021}.
On the other hand, studies of the isotopic composition of the Solar System bodies imply that Jupiter's core of $10-20~M_{\oplus}$ accreted within $<0.5~$Myr, while Jupiter reached $\sim 50~M_{\oplus}$ after $\sim 2~$Myr, and its final size of $\sim 318~M_{\oplus}$ before $\sim 4-5~$Myr \citep{Kruijer_et_al_2020}.
Therefore, instead of a fully-grown Jupiter, it is more reasonable and realistic to assume a younger and smaller Jupiter, which perturbs the inner disk while the terrestrial planets start to grow via planetesimal accretion. 
Meanwhile, to generalize the model to exoplanetary systems, considering a larger planet mass than Jupiter mass is also necessary.
Therefore, we consider four values of the mass of the (proto-) CJ: $M_{\rm{p}} = 0.1$, 0.3, 1, 3$~M_{\rm{J}}$. 

\subsubsection{planet eccentricity}
If we assume that Jupiter has formed by core accretion, its eccentricity might also deviate from the present value. 
Since the eccentricity of Jupiter oscillates periodically as a result of secular interaction with other planets in the Solar System \citep[e.g.,][]{Brouwer_vanWoerkom_1950}, the present value of 0.048 might not be appropriate if we consider the early stage before the accumulation of inner planets. 
After Jupiter's core becomes massive enough, it starts runaway gas accretion and the orbital eccentricity of the growing Jupiter is determined by the angular momentum of the accreted gas. 
\citet{Kikuchi_2014_apj} showed that the orbits of giant planet cores are circularized through the accretion of disk gas in the outer disk regions.
However, since we are unclear of the exact moments when Jupiter's core starts its rapid gas accretion and when the accretion of inner planets starts in the inner disk, we also vary the eccentricity of the planet to reveal its impact on the dynamics of inner planetesimals.
This also helps to generalize our results to exoplanetary systems harboring CJs of various eccentricities.
In accordance with the mass range, we also consider four values of planet eccentricity: $e_{\rm{p}} = 0.01$, 0.02, 0.05, 0.1.

\subsubsection{planet semi-major axis}

Recent studies suggest that CJs have very likely undergone orbital migration as a result of interaction with the protoplanetary disk or planetesimal disk. 
The Nice model \citep[e.g.,][]{Nice_model} predicts that the four giant planets in the solar system migrated because of the exchange of angular momentum with the planetesimal disk after the dissipation of nebula gas.
During an earlier stage where the gaseous disk is still present, the gas giants (e.g., Jupiter and Saturn) could have undergone significant inward and outward migration, shaping the structure of the inner Solar System. \citep{Walsh2011}.
Therefore, in the context of this study, it is necessary to consider the effect of orbital migration of the CJ. 
Simulating the migration of the planet would introduce other variables that control the migration as well as gas accretion (e.g., disk viscosity, opacity, etc.) and thus expand the parameter space.
For the sake of simplicity and demonstrating the effect of orbital alignment more clearly, we do not implement planet migration in our simulations but use different models with different values of the semi-major axis of the planet to investigate the dependence on this parameter. 

The Grand Tack model \citep{Walsh2011} assumes that a fully-formed Jupiter starts at 3.5 au, which is a location favorable for giant planet formation because of the presence of the snow line, and then it migrates inward to $\sim 1.5~$au, where it starts outward migration together with a growing Saturn. 
Note that the model in the Grand Tack scenario is different from that in this study. 
Firstly, Saturn is absent in our models, so that the condition for "tacking" of the CJ is not met.
In addition, we focus on demonstrating the effect of the orbital alignment of planetesimals under the secular perturbations from the CJ and gas drag, so we avoid placing the CJ too close to the planetesimals. 
If the CJ is located too close to the planetesimals ($a_{\rm{p}} \simeq 1.5~$au), most of the planetesimals would be scattered; meanwhile, the change of gas surface density due to the gap opening of the gas giant would also affect the dynamics of the planetesimals.
Therefore, we consider a minimum semi-major axis of the planet of 3.5 au.
Also, following KW00 and to account for possible outward migration of the planet, we consider a maximum semi-major axis of 6.2 au.
In total, we consider three values of planet semi-major axis: $a_{\rm{p}} = 3.5$, 5.2, and 6.2 au.
Changing the location of the planet also allows us to extrapolate our results to exoplanetary systems harboring CJs of various orbital distances.

\subsubsection{Models}

We perform in total 20 simulations as summarized in Table \ref{table:model_names}.
In models M1-4-S, E1-3-S, and A1-2-S, we investigate the dependence of the relative velocities of small planetesimals on the mass, eccentricity, and location of an outer planet.
In the meantime, by considering a mass (size) spectrum of the planetesimals, we show the size-dependent effect of orbital alignment on the relative velocities.
The model in which the planet parameters are equal to those of Jupiter, i.e., $M_{\rm{p}} = M_{\rm{J}}$, $e_{\rm{p}} = 0.05$, and $a_{\rm{p}} = 5.2~$au, is considered as the fiducial model.
We change one parameter when fixing the other two parameters as the fiducial values in order to see the dependence of the relative velocities of planetesimals on the varied planet parameter.
In models M1-4-S, E1-3-S, and A1-2-S, the planetesimals are treated as test particles, i.e., the mutual interaction of planetesimals is neglected.
Each model is integrated for approximately 30,000 years.

In models M1-4-L, E1-3-L, A1-2-L, E0, and M0, we perform full \textit{N}-body simulations and integrate the orbits of initially large planetesimals ($m=10^{23}~$g) near 1 au while varying the parameters of the outer planet.
Such a large initial size of the planetesimals is suggested by the streaming instability mechanism \citep[e.g.,][]{Youdin_2005}.
The typical size of planetesimals produced by streaming instabilities is approximately a few hundred kilometers \citep[e.g.,][]{Johansen_2012}.
Therefore, in addition to the "classical" small planetesimals as considered in previous studies, we adopt a uniform planetesimal mass of $10^{23}~$g (a radius of roughly 200 km when the bulk density $\rho = 3~\rm{g}~{cm}^{-3}$).
For such large planetesimals, the mutual gravitational interaction cannot be safely neglected. 
E0 and M0 are two control models demonstrating (for comparison) the influence of a Jupiter-mass planet on a circular orbit on the dynamics of inner planetesimals and the unperturbed case where no planet exists in the outer disk.
Each model is integrated for approximately 25,000 years.
These simulations are accelerated by using GPUs.

\begin{deluxetable*}{L|ccccccccc|ccccc}
\label{table:model_names}
\tablecaption{Models}
\tablehead{
\colhead{Planet parameters} & \colhead{M1} & \colhead{M2} & \colhead{M3} & \colhead{M4} & \colhead{E1} & \colhead{E2} & \colhead{E3} & \colhead{A1} & \colhead{A2} & \colhead{E0} & \colhead{M0}
}
\startdata
M_{\rm{p}}~(M_{\rm{J}}) & 0.1 & 0.3 & 1 & 3 & 1 & 1 & 1 & 1 & 1 & 1 & 0\\
e_{\rm{p}} & 0.05 & 0.05 & 0.05 & 0.05 & 0.01 & 0.02 & 0.1 & 0.05 & 0.05 & 0 & -\\
a_{\rm{p}}~(\rm{au}) & \multicolumn{7}{c}{5.2} & 3.5 & 6.2 & \multicolumn{2}{c}{5.2} \\
m~(\rm{g}) & \multicolumn{9}{c|}{$10^{16}$-$10^{20}$ (-S), $10^{23}$ (-L)} & \multicolumn{2}{c}{$10^{23}$}\\
\enddata
\tablecomments{The parameters of the planet (mass $M_{\rm{p}}$, eccentricity $e_{\rm{p}}$, semi-major axis $a_{\rm{p}}$) and mass of planetesimals $m$ used in each model. For models M1-4, E1-3, and A1-2, we performed two versions of simulations, one for the small planetesimal case ($m=10^{16-20}$~g, with a postfix "-S"), and one for the large planetesimal case ($m=10^{23}~$g, with a postfix "-L").}
\end{deluxetable*}

\subsection{Gas component} \label{subsec:gas_disk}

In our models, each system consists of a central star of mass $M_* = 1\, M_{\odot}$, a (proto-)CJ (except for model M11), and a protoplanetary disk comprised of gas and planetesimals.
The gas component of the protoplanetary disk follows a power law distribution:
\begin{equation}
    \Sigma_{\rm{g}} = 2.4 \times 10^3 \left(\frac{a}{\rm{au}} \right)^{-3/2} \rm{g}\, \rm{cm}^{-2},
    \label{eq:gas_profile}
\end{equation}
which corresponds to a 50\% more massive disk than the minimum mass solar nebula (MMSN) \citep{Hayashi_1981}.
The spatial density of gas at the disk midplane is 
\begin{equation}
    \rho_{\rm{g}} = 2.0 \times 10^{-9} \left( \frac{a}{1\,{\rm{au}}}\right)^{-11/4} \rm{g}\, \rm{cm}^{-2}.
    \label{eq:gas_density}
\end{equation}

Since we assume an axisymmetric circular gas disk, $a$ in the above equations is the distance from the central star.
Our simplified disk model might not be accurate enough to account for the true behavior of the gas disk in response to the gravitational perturbation of a giant planet. 
The gas disk can depart from perfectly circular streamlines when perturbed by a stellar companion \citep[e.g.,][]{Artymowicz1994}.
Apart from that, it can also become eccentric if a giant planet is embedded within, even if the planet is on a circular orbit \citep{Hsieh_Gu_2014}.
Previous studies report that the eccentricity of a circumprimary gas disk can be excited by a stellar companion and raise the relative velocities of planetesimals within by a factor of $\sim 2$ compared to the circular disk case \citep{Paardekooper2008}.
However, due to the uncertainties of the realistic behavior of the inner gas disk under the perturbation of a CJ, we use the simplified circular disk model as a first step to clarify the basic physics dominating the dynamics of planetesimals in such a system configuration. 
We also neglect the possible gap opened by the CJ in the gaseous disk because we mainly focus on the orbital evolution of planetesimals in the inner disk region far away from the location of the planet.

\subsection{Planetesimals}

In models M1-4-S, E1-3-S, and A1-2-S, we consider a mass spectrum of planetesimals: $m = [10^{16}, 10^{17}, 10^{18}, 10^{19}, 10^{20}]~\rm{g}$.
For each planetesimal mass, we have $N = 10,000$ particles initially distributed in an annulus of 1-4 au, following a power law surface density distribution $\Sigma_{\rm{d}} \propto a^{-3/2}$, with an exception of model A1-S, in which the planetesimals are initially placed within 1-2.7 au, with the same solid surface density profiles as those in other models (M1-4-S, E1-3-S, and A2-S), in order to avoid close encounters with the planet located at 3.5 au. 
The initial eccentricity and inclination of planetesimals follow the Rayleigh distribution with dispersion $\sigma_e = \sigma_i = 2 r_{\rm{H}}/a$, where the Hill radius of the planetesimal $r_{\rm{H}}$ given by Eq. (4) in GK21.
The other orbital elements $(\Omega, \omega, \tau)$ are randomly chosen from 0 to $2\pi$.
The bulk density of the planetesimals is $\rho = 3~\rm{g}~cm^{-3}$.

In models M1-4-L, E1-3-L, A1-2-L, E0, and M0, the planetesimals have a uniform mass of $10^{23}~$g and are initially distributed in a thin ring near 1 au.
The solid surface density is similar to that of the MMSN \citep{Hayashi_1981}
\begin{equation}
\Sigma_{\rm{d}} = 10 \times \left(\frac{a}{1~\rm{au}}\right)^{-3/2}~\rm{g}~cm^{-2}.
\label{eq:MMSN_Sigma_d}
\end{equation}
The number of planetesimals is $N = 2048$ in each model. 
Since viscous stirring is important for planetesimals of such large mass, the initial dispersion of eccentricity and inclination is given by the equilibrium values under gas damping and viscous stirring \citep{Kokubo_Ida_2000}

\begin{eqnarray}
    e_{\rm{eq}} = 2 i_{\rm{q}} \simeq &4.3& \times 10^{-3} \left(\frac{m}{10^{23}~\rm{g}}\right)^{4/15} \left(\frac{\Sigma_{\rm{d}}}{10~\rm{g}~cm^{-2}}\right)^{1/5} \nonumber\\
    &\times& \left(\frac{\rho_{\rm{g}}}{2\times 10^{-9}~\rm{g}~cm^{-3}}\right)^{-1/5} \left(\frac{a}{1~\rm{au}}\right)^{1/5}.
    \label{eq:e_eq}
\end{eqnarray}

The planetesimals feel the gas drag which is calculated by \citep{Adachi_et_al_1976}:
\begin{equation}
    f_{\rm{g}} = \frac{1}{2}C_{\rm{D}} \pi s^2 \rho_{\rm{g}} u^2
    \label{eq:gas_drag}
\end{equation}
where $C_{\rm{D}}$ is the non-dimensional drag coefficient, $s$ is the particle radius, and $u$ is the relative velocity between the gas and the particle.
The value of $C_{\rm{D}}$ is described in Section 2.1 in GK21.
The timescale of gas damping effect on planetesimal eccentricity is given by \citet{kokubo2012dynamics}
\begin{eqnarray}
    \tau_{\rm{damp}} = \frac{e}{|{\rm{d}}e/{\rm{d}}t|} \sim 10^5 \left(\frac{e}{0.1}\right)^{-1} \left(\frac{m}{M_{\oplus}}\right)^{1/3} \nonumber\\
    \left(\frac{\rho}{3\, {\rm{g}}\,{\rm{cm}}^{-3}}\right)^{2/3} \left(\frac{a}{1\,{\rm{au}}} \right)^{13/4}\, \rm{yr}.
    \label{eq:tau_damp}
\end{eqnarray}

The equation of motion of a planetesimal is then 
\begin{eqnarray}
\odv{\boldsymbol{v}_i}{t} = &-& GM_* \frac{\boldsymbol{x}_i}{|\boldsymbol{x}_i|^3} + G M_{\rm{p}} \frac{\boldsymbol{x}_{\rm{p}}-\boldsymbol{x}_i}{|\boldsymbol{x}_{\rm{p}} - \boldsymbol{x}_i|^3} \nonumber \\ 
&-& GM_{\rm{p}} \frac{\boldsymbol{x}_{\rm{p}}}{|\boldsymbol{x}_{\rm{p}}|^3} + \boldsymbol{f}_{\rm{g}} + \sum_{j\neq i}^N Gm_j \frac{\boldsymbol{x}_j-\boldsymbol{x}_i}{|\boldsymbol{x}_j-\boldsymbol{x}_i|^3}.
\label{eq:eom}
\end{eqnarray}
The subscript ``p'' denotes the planet parameters.
The third term is the indirect term, which arises from the indirect perturbation of the planet on the central star.
The last term describes the mutual gravitational interaction of planetesimals and is only included in models M1-4-L, E1-3-L, A1-2-L, E0, and M0.

\subsection{Relative velocity}  \label{subsec:method_velocity}

The relative velocity of planetesimals is calculated using the method described in \citet{Guo_2022} (hereafter GK22). 
In order to give more general results, in models considering small planetesimals ("-S"), we average the values of the relative velocity $\langle \Delta v \rangle$ over a time span from $t_{\rm{start}} \simeq 2~$Myr to $t_{\rm{end}} \simeq 3~$Myr to avoid the initial small values.
In models considering large planetesimals ("-L", E0, and M0), the time evolution as well as the time-averaged values of the relative velocities (velocity dispersion) are demonstrated.
For more details on how to calculate $\langle \Delta v \rangle$, we refer to GK22.

\section{Results} \label{sec:results}

The results of the models in which the planetesimals are treated as test particles (small planetesimal case) are shown first.
The dependence of the relative velocities of planetesimals on the planet parameters and mass ratio of the planetesimals is shown in section \ref{subsec:M1-M7}.
Following that, we show the results of models considering initially large planetesimals and including their mutual gravitational interaction in section \ref{subsec:new_results}.

\subsection{The Small Planetesimal Case} \label{subsec:M1-M7}

\subsubsection{Time evolution} \label{subsec:time_evolution}

In order to understand the dynamical behavior of planetesimals under the coupling effect of secular perturbation and gas drag, we first estimate the timescales of these effects separately. 
The timescale of orbital decay is also calculated in order to check the inward drift of planetesimals. 

The timescale of orbital decay is given by 
\begin{equation}
    \tau_{\rm{decay}} = \frac{a}{{\rm{d}}a/{\rm{d}}t},
\end{equation}
where the decay rate of the semi-major axis is given by \citet{Adachi_et_al_1976}
\begin{eqnarray}
    \odv{a}{t} &\simeq& -1.3 \times 10^{-2} \left( \frac{5}{8} e^2 + \frac{1}{2} i^2 + \eta^2\right)^{1/2} \left( e^2 + \frac{1}{8} i^2 + \eta \right) \nonumber \\ 
    &\times& \left(\frac{m}{10^{23}~{\rm{g}}}\right)^{-1/3} \left(\frac{\rho_{\rm{gas}}}{2\times 10^{-9}~ {\rm{g}}~ {\rm{cm}}^{-3}}\right) \nonumber \\
    &\times& \left(\frac{a}{1~ {\rm{au}}}\right)^{1/2} ~ ({\rm{au}}~ {\rm{yr}}^{-1}).
\label{eq:tau_decay}
\end{eqnarray}
Here $\eta$ is a small quantity to account for the lag between the rotation velocity of the gas component and the local Keplerian velocity $v_{\rm{K}}$. 
For our model setup, $\eta$ is given by \citet{Kokubo_Ida_2000}
\begin{equation}
    \eta = 0.0019\left(\frac{a}{1\, {\rm{au}}}\right)^{1/2}.
\end{equation}
For the eccentricity $e$ in Eq. (\ref{eq:tau_decay}), we use the time-averaged value of the mean eccentricity of particles in the given bin of the semi-major axis. 
The values are averaged over $\simeq 30000$ years.

The timescale of gas damping is given in Eq. (\ref{eq:tau_damp}).
The timescale of secular perturbation $\tau_{\rm{sec}}$ is
\begin{equation}
    \tau_{\rm{sec}} = \frac{2\pi}{A} = \frac{8 \pi a_{\rm{p}}^3 n \beta^3}{3GM_{\rm{p}}} = \frac{8 \pi n}{3 \mu n_{\rm{p}}^2}(1-e_{\rm{p}}^2)^{3/2},
    \label{eq:tau_sec}
\end{equation}
where $n$ is the mean motion of the planetesimal, $\beta = \sqrt{1-e_{\rm{p}}^2}$, and the coefficients $A$ and $B$ (used in Eq. \ref{eq:forced_eccentricity}) are
\begin{eqnarray}
    A &=& \mu \frac{3}{4} \frac{n_{\rm{p}}^2}{n}\frac{1}{\beta^3}, \label{eq:A}\\
    B &=& \mu \frac{15}{16} \frac{n_{\rm{p}}^2}{n} \frac{a}{a_{\rm{p}}} \frac{e_{\rm{p}}}{\beta^5}.
    \label{eq:B}
\end{eqnarray}
Here $\mu$ is the mass ratio $\mu = M_{\rm{p}}/(M_{\rm{p}}+M_*)$.

\begin{figure}[ht]
    \includegraphics[width=0.45 \textwidth]{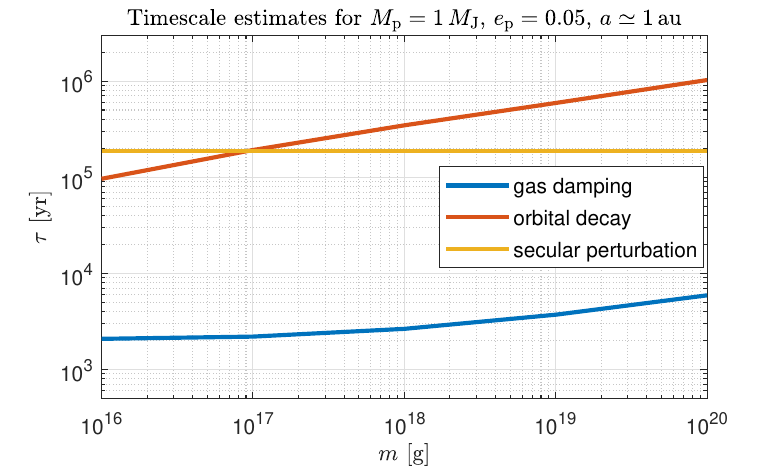}
    \caption{Timescales of gas damping, orbital decay, and secular perturbation near 1 au for the fiducial model. The timescale of secular perturbation is independent of the particle mass.}
    \label{fig:inner_timescales}
\end{figure}
    
Fig. \ref{fig:inner_timescales} shows the timescale estimates at $a\simeq 1\,{\rm{au}}$, which corresponds to the typical terrestrial planet forming region. 
The timescale of orbital decay is much longer than our simulation time span ($\simeq 30000$ years), so that the particles in our model do not suffer from severe inward drift.
The short timescale of gas damping indicates that the eccentricities of the particles reach equilibrium values quickly (within $\sim \mathcal{O}(10^3)$ years). 
Although we are only showing the timescale estimates for the fiducial model for example, the large distinction between the gas damping timescale and the other two timescales holds for all the other models as well. 
In other words, the gas-damping timescale is always much shorter than the timescales of orbital decay and secular perturbation in models concerning small planetesimals.


\subsubsection{Spatial distribution of orbital elements} \label{subsec:inner_orbit}

\begin{figure*}
    \includegraphics[width= \textwidth]{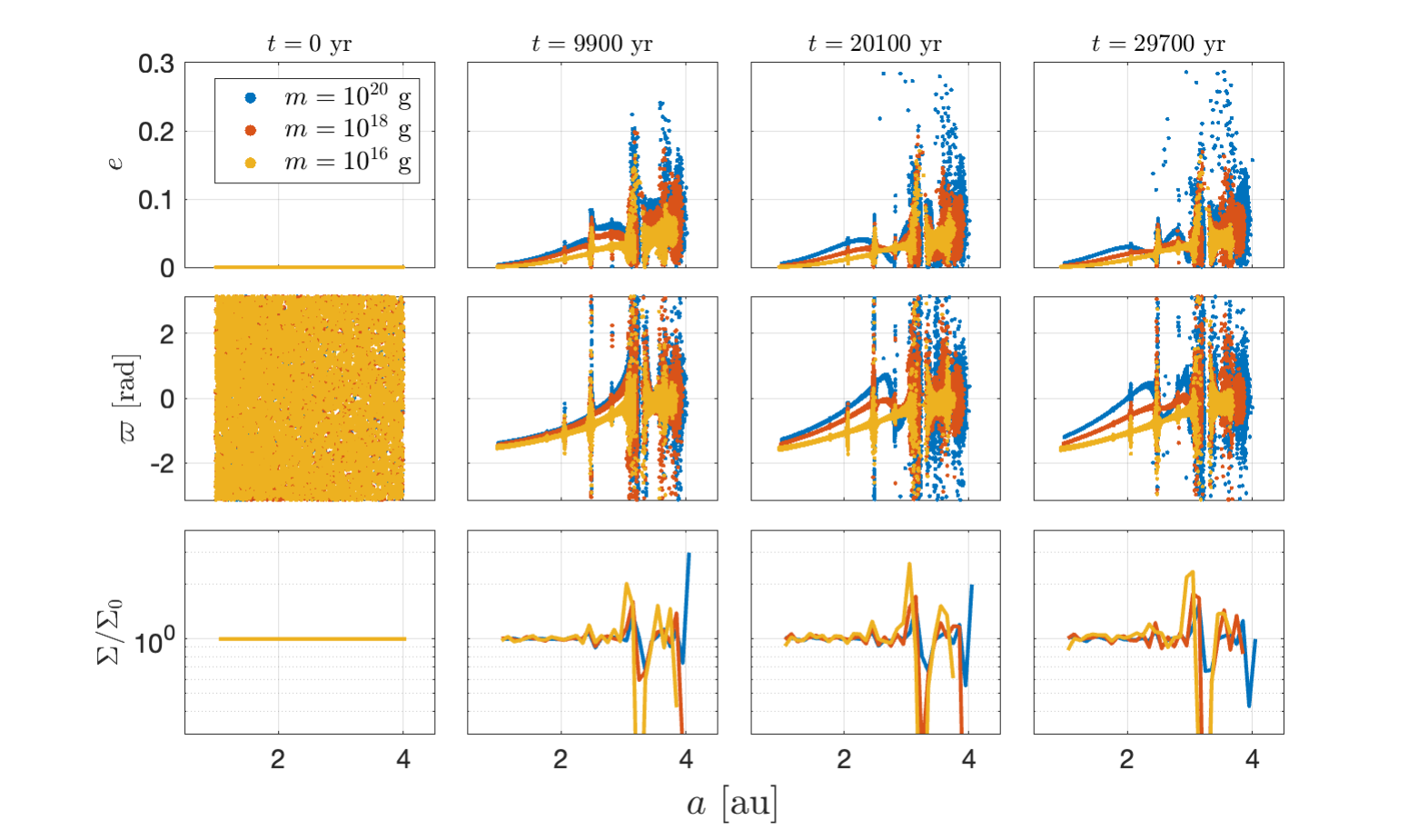}
    \caption{Distribution of the eccentricity (first row), longitude of pericenter (second row), and solid surface density (third row) along the semi-major axis. The four columns represent four equally-spaced (roughly) times of evolution: $t = 0$, $t \simeq 10000$, $t \simeq 20000$, and $t \simeq 30000$ years. The colors of the dots represent the mass of the particles.}
    \label{fig:spatial_distribution_inner}
\end{figure*}

The distribution of the eccentricity, longitude of pericenter, and solid surface density along the semi-major axis is shown in Fig. \ref{fig:spatial_distribution_inner}.
The snapshots are taken at four equally-spaced times $t \simeq 0$, 10000, 20000, and 30000 years to capture the time evolution.
The location of the 3:1 and 2:1 mean motion resonance (MMR) can be seen at $a\simeq 2.5$ au and $a\simeq 3.3$ au, respectively. 
At the MMRs, the eccentricities are excited and deviate from the average values.
Meanwhile, the longitudes of pericenter are scattered, demonstrating that the orbits are not aligned at these locations. 
As a result of scattering at MMRs and subsequent gas damping, the particles temporarily pile up at the inner edge of the MMRs and leave a dip of solid surface density on the outside. 
At non-MMR locations, the eccentricities are generally low and the longitudes of pericenter are relatively well-aligned.
In the disk region where the orbits are rather "chaotic" (outside the 2:1 MMR), the smaller-mass particles have comparatively better-aligned orbits, while the eccentricities and longitudes of pericenter of large-mass particles are scattered in a large range. 
At 1 au, the orbits of all particle masses are always well aligned.

\begin{figure}
    \centering
    \includegraphics[width=0.45\textwidth]{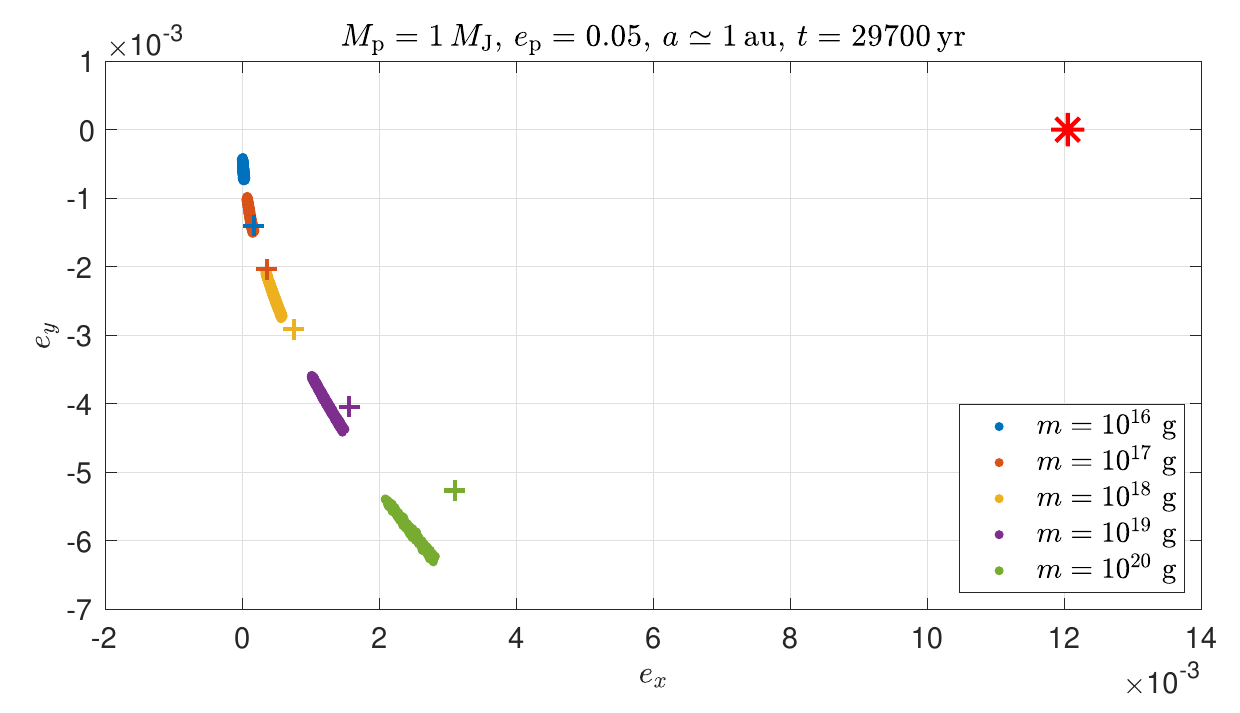}
    \caption{Distribution of eccentricity vectors near 1 au. The red asterisk marks the forced eccentricity induced by the planet. The "+" signs show the equilibrium locus of the eccentricity vectors. Different colors are for different particle masses as indicated in the legend. The width of the semi-major axis bin is $\Delta a = 0.1~$au.}
    \label{fig:eccvec_1au}
\end{figure}

The alignment of orbits can be visualized on the eccentricity vector ($e_x$-$e_y$) plane more clearly. 
Fig. \ref{fig:eccvec_1au} shows the distribution of the eccentricity vector of particles near 1 au for the fiducial model at $t = 29700~$ year (the end of the simulation). 
The forced eccentricity $e_{\rm{f}}$ imposed by the perturber is given by 
\begin{equation}
    e_{\rm{f}} = \frac{B}{A} = \frac{5}{4} \frac{a}{a_{\rm{p}}} \frac{e_{\rm{p}}}{\beta^2}.
    \label{eq:forced_eccentricity}
\end{equation}
The elongated distribution (especially for larger-mass populations) is due to the differential precession within the finite width of the semi-major axis bin ($\Delta a = 0.1~$au). 
The eccentricity vectors of particle populations with different masses (marked by different colors) are separated on the $e_x$-$e_y$ plane by their distinct response to gas drag.
The equilibrium locus of eccentricity vectors are also plotted for reference. 
Further discussion on the equilibrium locus of the eccentricity vector can be found in Section 3.2 in GK21. 
The distribution of eccentricity vectors of each particle mass is very close to the corresponding equilibrium locus, suggesting that the eccentricity vectors have reached (or almost reached) equilibrium under the coupling effect of secular perturbation and gas damping.
For the largest mass ($m=10^{20}\,$g), the eccentricity vectors are still evolving towards the equilibrium location. 
It is clearly shown that the orbits of equal-mass particles near 1 au are well aligned - a feature consistent with what was found in KW00.

\subsubsection{Relative velocities} \label{subsec:velocity}

\begin{figure}
    \centering
    \includegraphics[width=0.45\textwidth]{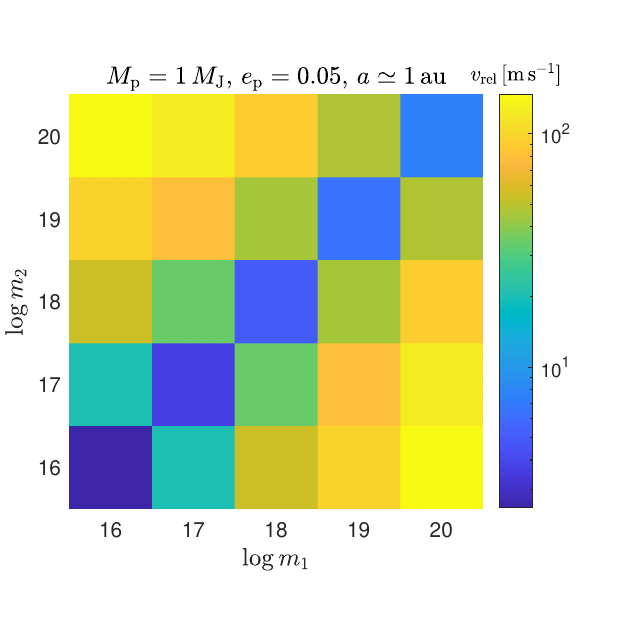}
    \caption{Relative velocity between particles of $m_1$ and $m_2$ near 1 au for the fiducial model. The velocities are averaged over $\simeq 20000$ years; the values from the first $\simeq 10000$ years are skipped to avoid the initially small values due to the initial conditions.}
    \label{fig:v_rel_m3e3_1au}
\end{figure}

Before showing the results of our dependence study, we compare our results with those of KW00.
Fig. \ref{fig:v_rel_m3e3_1au} is a color map showing the relative velocity $\langle \Delta v \rangle$ between particles of masses $m_1$ and $m_2$ near 1 au. 
We can identify two characteristic trends: (i) the relative velocities of equal-mass particles (i.e. the velocity dispersion $\sigma_v$) remain very low and increase with increasing particle mass; (ii) the relative velocities between different-mass particles increase as the mass ratio of $m_1$ and $m_2$ increases. 
Near 1 au, $\sigma_v$ is on the order of 1 m$\, \rm{s}^{-1}$, which is consistent with the result in KW00.

We demonstrate the dependence of the relative velocities on the planet parameters by focusing on the relative velocities of particles near 1 au.

\vspace{12pt}
\noindent \textit{Dependence on planet mass}  
\vspace{4pt}

Changing the mass of the planet $M_{\rm{p}}$ has a straightforward influence on the relative velocities of planetesimals through its impact on the degree of alignment of planetesimal orbits. 
Eq. (\ref{eq:forced_eccentricity}) shows that $e_{\rm{f}}$ does not depend on $M_{\rm{p}}$.
With $e_{\rm{f}}$ fixed, the impact of the increasing planet mass is mainly reflected by the varying timescale of secular perturbation. 
The timescale of secular perturbation $\tau_{\rm{sec}}$ decreases by over one order of magnitude as $M_{\rm{p}}$ increases from 0.1 to 3$\, M_{\rm{J}}$. 
As a consequence, the eccentricity vectors experience more differential precession in a given range of semi-major axis on the $e_x$-$e_y$ plane.
The more sheared orbits lead to higher relative velocities at specific locations.
Another effect of increasing planet mass is that the short-term gravitational kick during each encounter with the planet becomes stronger. 
This also results in larger dispersion of the eccentricity vectors at a specific semi-major axis. 

\begin{figure}
    \centering
    \includegraphics[width=0.45 \textwidth]{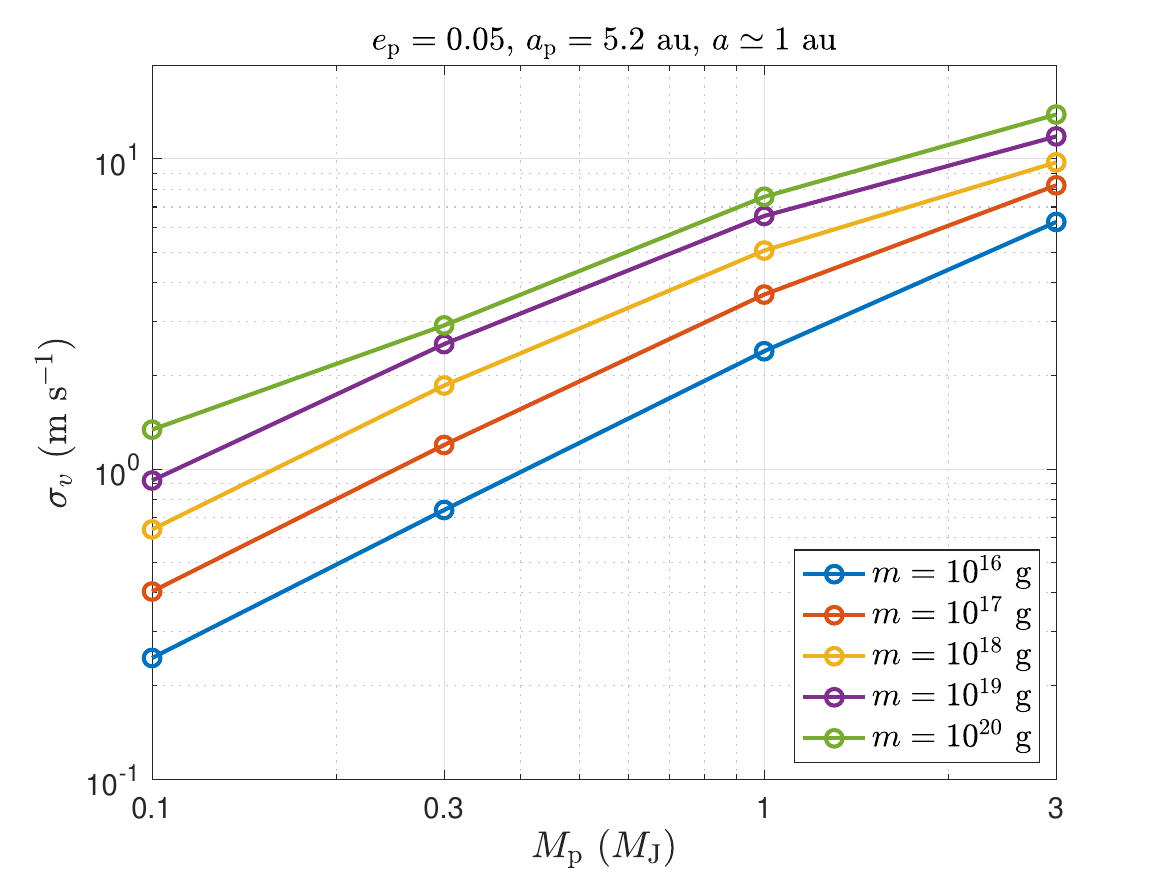}
    \caption{Dependence of $\sigma_v$ on planet mass $M_{\rm{p}}$ near 1 au. The $\sigma_v$ values are averaged over a time span starting from $\simeq 10,000$ yrs after the beginning till the end of the simulation ($\simeq 20,000$ yrs in total).}
    \label{fig:sigmaV_Mp_1au}
\end{figure}

Therefore, the variation of the planet mass affects $\langle \Delta v \rangle$ on both the long term and the short term: the secular perturbation timescale $\tau_{\rm{sec}}$ and the gravitational kick during each encounter (GK22). 
Fig. \ref{fig:sigmaV_Mp_1au} shows the dependence of the velocity dispersion among equal-mass particles $\sigma_v$ on planet mass $M_{\rm{p}}$. The values of $\sigma_v$ here are averaged over a time span of roughly 20,000 years ($t \simeq 10000$-30000 yr; the results from the first 10,000 years are neglected in order to avoid the initially small values set in the initial conditions).
For all particle masses, $\sigma_v$ is always a monotonically increasing function of $M_{\rm{p}}$.
The dependence of the relative velocities on both the planet mass and the particle mass ratio is summarized in Fig. \ref{fig:v_compare_1au}.
It shows that $\langle \Delta v \rangle$ increases monotonically with increasing mass ratio and increasing $M_{\rm{p}}$.

\vspace{12pt}
\noindent \textit{Dependence on planet eccentricity}
 \vspace{4pt}

Similarly, increasing planet eccentricity $e_{\rm{p}}$ also leads to a larger dispersion of the eccentricity vectors and higher relative velocities of planetesimals. 
Eq. (\ref{eq:tau_sec}) shows that $\tau_{\rm{sec}}$ decreases as $e_{\rm{p}}$ increases. 
However, the dependence of $\tau_{\rm{sec}}$ on $e_{\rm{p}}$ is very weak compared to that on $M_{\rm{p}}$.
Therefore, the impact of increasing $e_{\rm{p}}$ on the degree of orbital alignment is realized by its influence on the forced eccentricity.
As discussed in Section 3 in GK21, the eccentricity vector $(k,h)$ of a particle under secular perturbation circulates the forced eccentricity $(e_{\rm{f}} \cos{\varpi_{\rm{p}}}, e_{\rm{f}} \sin{\varpi_{\rm{p}}})$ with angular velocity $A$ and radius $e_{\rm{f}}$.
Neglecting the correction of the eccentricity vector induced by gas drag (since it is quite small compared to the magnitude of $e_{\rm{f}}$), for almost the same $\tau_{\rm{sec}}$, larger magnitude of $e_{\rm{f}}$ leads to longer arc of circulation, which means the distribution of eccentricity vectors are more spread-out on the $e_x$-$e_y$ plane.
Therefore, the orbits of particles near 1 au becomes less aligned as the planet eccentricity increases. 
 
\begin{figure}
    \centering
    \includegraphics[width=0.45 \textwidth]{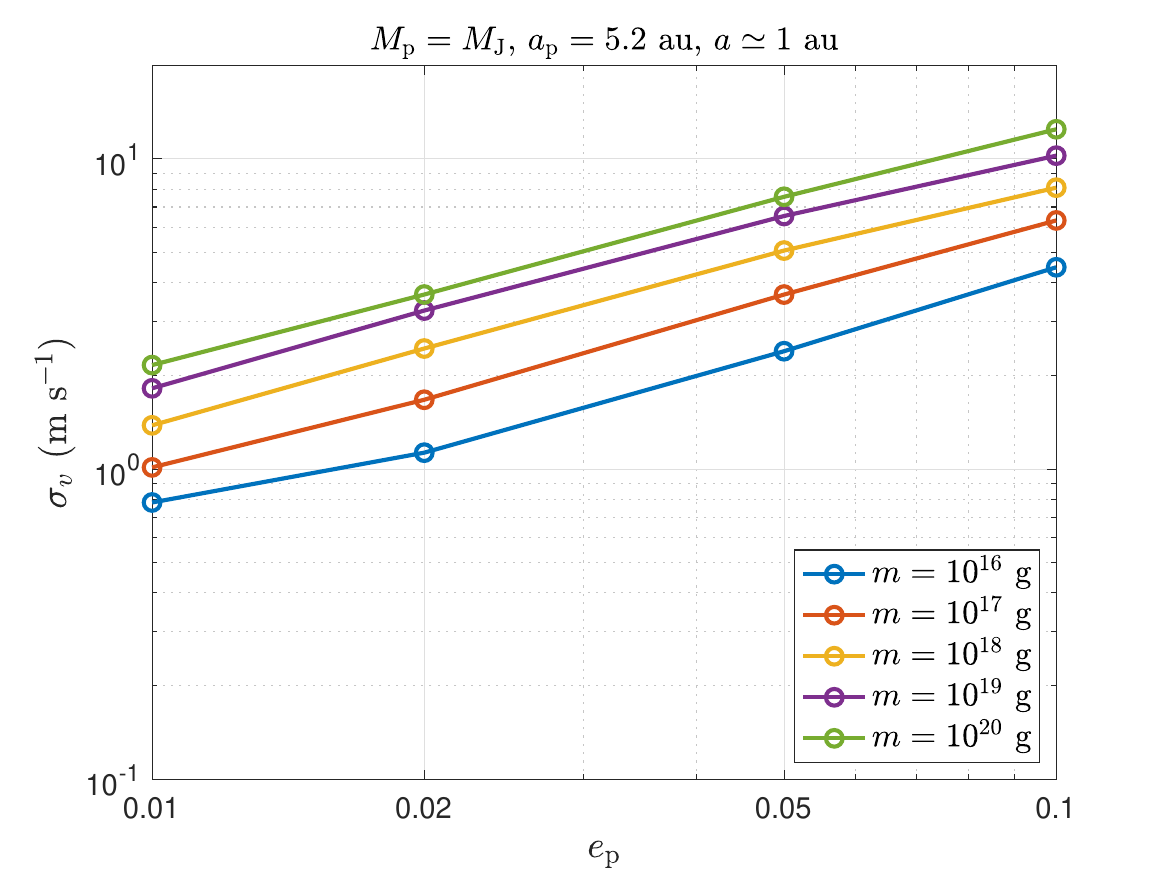}
    \caption{Similar to Fig. \ref{fig:sigmaV_Mp_1au}, but shows the dependence of $\sigma_v$ on planet eccentricity $e_{\rm{p}}$ near 1 au.}
    \label{fig:sigmaV_ep_1au}
\end{figure}

Qualitatively, the dependence of planetesimal velocities on planet eccentricity is the same as their dependence on planet mass, i.e., the values of $\langle \Delta v \rangle$ of all planetesimal masses increase monotonically with increasing $e_{\rm{p}}$, as shown in Fig. \ref{fig:sigmaV_ep_1au}.  
However, the dependence is weaker than that on the planet mass, as shown by the flatter slope in Fig. \ref{fig:sigmaV_ep_1au} compared to that in Fig. \ref{fig:sigmaV_Mp_1au}.
Such a dependence is a result of the change in the forced eccentricities $e_{\rm{f}}$ as we vary the planet eccentricity $e_{\rm{p}}$, which leads to a more scattered distribution of the eccentricity vectors on the $e_x$-$e_y$ plane.

\vspace{12pt}    
\noindent \textit{Dependence on planet location} 
\vspace{4pt}

\begin{figure}
    \centering
    \includegraphics[width=0.45 \textwidth]{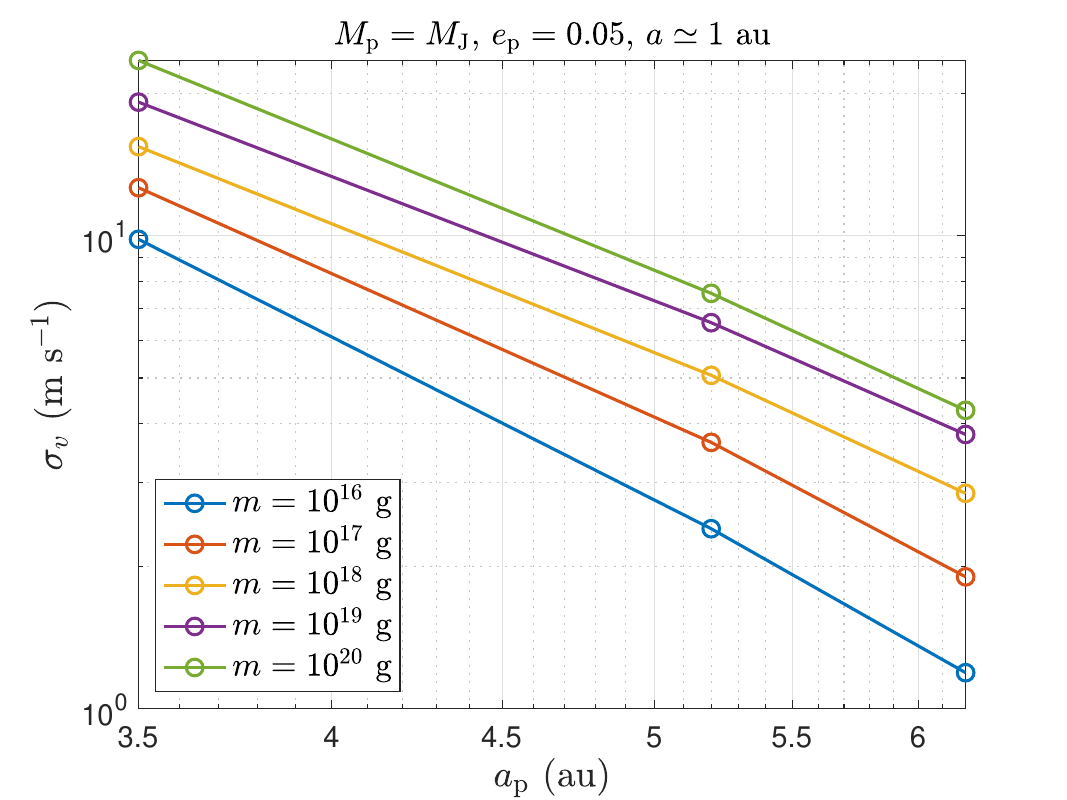}
    \caption{Similar to Fig. \ref{fig:sigmaV_Mp_1au}, but shows the dependence of $\sigma_v$ on planet semi-major axis $a_{\rm{p}}$ near 1 au.}
    \label{fig:sigma_v_ap_S}
\end{figure}

For small planetesimals ($m = 10^{16}-10^{20}~$g), the influence of reducing the semi-major axis of the planet is two-fold: it reduces the timescale of secular perturbation (see Eq. \ref{eq:tau_sec}) and increases the forced eccentricity (see Eq. \ref{eq:forced_eccentricity}).
Both of these effects result in more sufficient differential precession of the eccentricity vectors of the planetesimals and thus larger dispersion $\sigma_e$.
As discussed in section \ref{subsec:inner_orbit}, the time evolution of the eccentricity vectors of these small planetesimals near 1 au reach the equilibrium within a few thousand years.
Neglecting the mutual gravitational interaction of these planetesimals, the variation of the velocity dispersion due to varying planet location is dominated by the change of the strength of secular perturbation. 
Therefore, as $a_{\rm{p}}$ decreases, the secular perturbation becomes stronger, resulting in more sufficient shearing of orbits and thus higher relative velocities of a swarm of planetesimals. 
Fig. \ref{fig:sigma_v_ap_S} shows the dependence of the velocity dispersion of each planetesimal mass near 1 au on the semi-major axis of a Jupiter-like planet ($M_{\rm{p}} = M_{\rm{J}}$, $e_{\rm{p}} = 0.05$).
For all the planetesimal masses, $\sigma_v$ decreases with increasing $a_{\rm{p}}$.
Fig. \ref{fig:v_compare_1au} shows the dependence of the relative velocities between planetesimals with different masses near 1 au on the semi-major axis of the planet.
As shown in each panel, $\langle \Delta v \rangle$ is raised for all the planetesimal mass ratios when the planet is located on a smaller semi-major axis. 
These are the natural outcome of stronger secular perturbation.

\begin{figure*}
    \centering
    \includegraphics[width=0.93\textwidth]{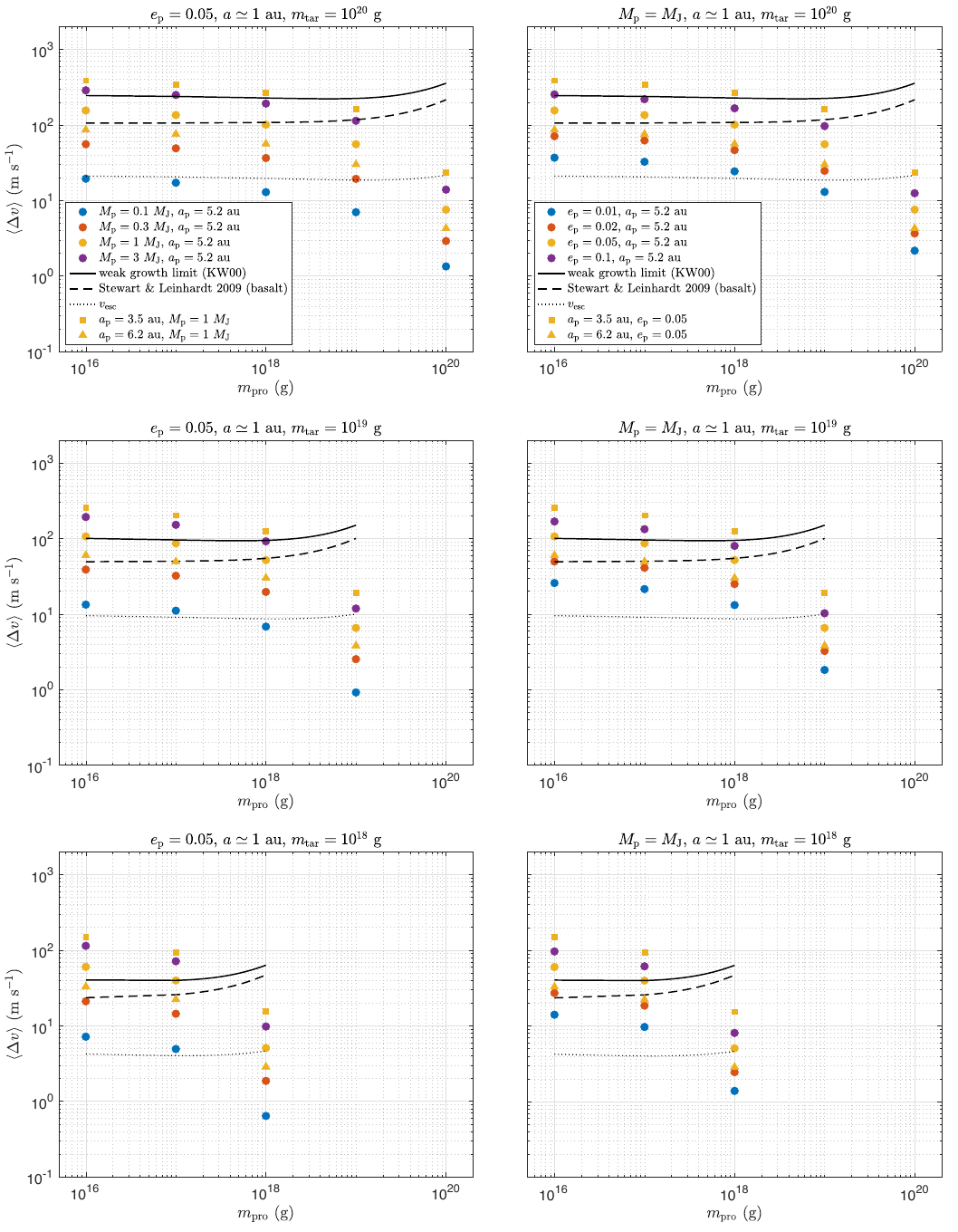}
    \caption{Relative velocities between particles of given mass ratios near 1 au. Each row corresponds to a fixed target mass. The subplots in the left column show the dependence of the relative velocities on the planet mass. Those in the right column show the dependence on the planet eccentricity. In both columns, the dependence on planet semi-major axis is shown (see legends). The solid curves represent the critical encounter velocity at the weak growth limit given in KW00 for reference. The dashed curves show the growth limit for basalt given in \citet{stewart2009velocity}. The dotted curves represent the two-body surface escape velocities $v_{\rm{esc}}$ given in Eq. (\ref{eq:escape_velocity}).}
    \label{fig:v_compare_1au}
\end{figure*}

\subsubsection{Growth limits near 1 au} \label{subsec:growth_limits}

The relative velocities not only depend on the mass ratio but also on the absolute value of planetesimal masses. 
In order to provide a complete picture of the dynamical and accretional behavior of the planetesimals, we compare the growth limits with the relative velocities between different-mass particles with varied target mass.
Since we are mainly interested in the formation of terrestrial planets, we focus on showing the relative velocities of planetesimals near 1 au.

Fig. \ref{fig:v_compare_1au} shows the relative velocities of planetesimals near 1 au in models considering small planetesimals ("-S").
The growth limits are also shown in comparison so that we can find the conditions that allow the accretion of planetesimals.
The definition of the growth limits is given in Section \ref{subsec:velocity}.
Here we calculate the growth limits with a particle density of $\rho = 3~\rm{g}~cm^{-3}$ since we focus on planetesimals near 1 au.
The two-body surface escape velocities are also plotted for reference. 
The escape velocity is given by
\begin{equation}
    v_{\rm{esc}} = \sqrt{\frac{2G(m_1+m_2)}{R_1+R_2}},
    \label{eq:escape_velocity}
\end{equation}
where $R_1$ and $R_2$ are the radius of the spherical particles.
Considering a more conservative growth limit given by \citet{stewart2009velocity}, planetesimal accretion is safe for the entire particle mass range near 1 au when ($e_{\rm{p}} = 0.05$, $M_{\rm{p}} \lesssim 0.3~M_{\rm{J}}$), ($e_{\rm{p}} \lesssim 0.02$, $M_{\rm{p}} = M_{\rm{J}}$), or when a Jupiter-like planet is located beyond $\simeq 6.2$ au.
A Jupiter-like planet at 5.2 au would prevent the growth of targets of masses $\lesssim 10^{18}~$g with a mass distribution (log of mass ratio $\geq 1$).
Planets larger than Jupiter would likely prevent the collisional growth of any bodies below $10^{20}~$g, other than equal-mass populations.
When $M_{\rm{p}} = M_{\rm{J}}$, and eccentricity larger than $\lesssim 0.05$ would likely suppress the growth of targets smaller than $\simeq 10^{19}~$g and log of mass ratio $\geq 1$.
When a Jupiter-like planet is located at $\simeq 3.5$ au, it is difficult for small planetesimals with a size distribution to grow via collisions. 
When a Jupiter-like planet is located further out at 6.2 au, it generally allows collisional growth of different-size bodies when the target mass is $\gtrsim 10^{19}~$g.
Runaway growth of bodies smaller than $\lesssim 10^{20}$ is typically allowed only for equal-mass bodies. 

The condition for accretion might be met for a wider range of parameters if we consider a looser growth limit (e.g., the weak growth limit given by KW00) or more massive planetesimals.

\subsection{The Large Planetesimal Case} \label{subsec:new_results}

Now we show the results of models M1-4-L, E1-3-L, A1-2-L, E0, and M0, where full \textit{N}-body simulations of planetesimals of a larger mass ($10^{23}~$g) are performed.

\subsubsection{Time evolution of eccentricities} \label{ssubsec:e_evo_m23}

\begin{figure*}
    \centering
    \includegraphics[width=0.99 \textwidth]{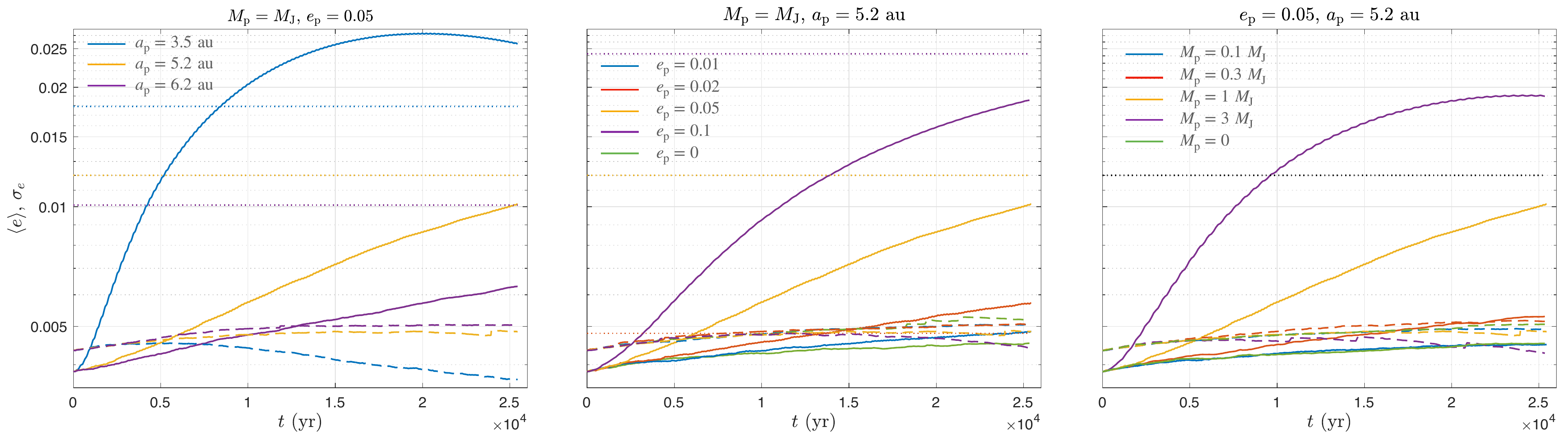}
    \caption{Time evolution of the mean (solid curves) and the dispersion (dashed curves) of the eccentricities of planetesimals near 1 au. Left: dependence on the location of the planet; middle: dependence on the eccentricity of the planet; right: dependence on the mass of the planet. The dotted lines indicate the forced eccentricities.}
    \label{fig:m23_e_evo}
\end{figure*}

For large planetesimals of $\sim 100~$km in size ($m=10^{23}~g$ in our simulations), the gas damping timescale is long and the equilibrium of the eccentricity vectors under gas damping, viscous stirring, and secular perturbation is not reached within a few tens of thousand years. 
Therefore, we first show the time evolution of the mean eccentricity $\langle e \rangle$ and the dispersion of the eccentricities $\sigma_e$ of the planetesimals.
Panels in Fig. \ref{fig:m23_e_evo} demonstrate the time evolution of the mean and dispersion of the eccentricities of planetesimals near 1 au and how these values depend on the location, eccentricity, and mass of the planet.
Regardless of the parameters of the planet, the mean eccentricity of the planetesimals increases with time as the eccentricity vectors move towards the forced eccentricity, which is an interplay of the secular perturbation from the planet and gas drag. 
Although we only integrate the planetesimal orbits for approximately 25,000 years, we can infer the subsequent evolution of $\langle e \rangle$: it would oscillate around the forced eccentricity with decreasing amplitude until equilibrium is reached. 
The mean eccentricities are raised to the highest value when the planet is located on the closest orbit (3.5 au), or when the planet eccentricity is the largest ($e_{\rm{p}} = 0.1$), since the forced eccentricity increases with decreasing $a_{\rm{p}}$ and increasing $e_{\rm{p}}$ (see Eq. \ref{eq:forced_eccentricity}).
Within the limited simulation time span, the mean eccentricities of the planetesimals also increase as the planet mass becomes larger owing to a reduced timescale of secular perturbation (see Eq. \ref{eq:tau_sec}), despite that the forced eccentricity remains unchanged as the planet mass varies.
As a result of increasing $\langle e \rangle$, the gas damping acting on these planetesimals becomes more efficient, so that the dispersion of eccentricity $\sigma_e$ becomes smaller.
Therefore, $\sigma_e$ reaches the lowest values when the planet is located on the closest orbit. 
For the same reason, a decrease of $\sigma_e$ is also observed when increasing the planet mass and eccentricity, but the amount of decrease is insignificant compared to that when reducing the planet semi-major axis. 

The model in which the planet is on a circular orbit ($e_{\rm{p}} = 0$) is similar to the model where there is no planet ($M_{\rm{p}}=0$).
When $e_{\rm{p}}=0$, the forced eccentricity is 0, so that there is no excitation of the mean eccentricity due to secular perturbation. 
The eccentricities of planetesimals evolve under the effect of gas drag and viscous stirring. 
Since the initial $\sigma_e$ is set to be the equilibrium value under gas drag and viscous stirring, $\sigma_e$ barely evolves in the two models.
In the case of a Jupiter-like planet (as well as a more massive or more eccentric planet), the mean eccentricity is raised by the secular perturbation towards the forced eccentricity, and therefore $\sigma_e$ slightly decreases as a result of more efficient gas damping.

When the large planetesimals are perturbed by a planet on an eccentric orbit, the differential precession of their $\boldsymbol{e}$ is suppressed mainly by the gravitational viscosity of the planetesimals (dynamical friction), so that when the strength of secular perturbation increases as the planet parameters change, $\sigma_e$ does not increase due to shear of orbits as in the small planetesimal cases.


\subsubsection{Dependence of random velocities on planet parameters}

\begin{figure*}
    \centering
    \includegraphics[width=0.99 \textwidth]{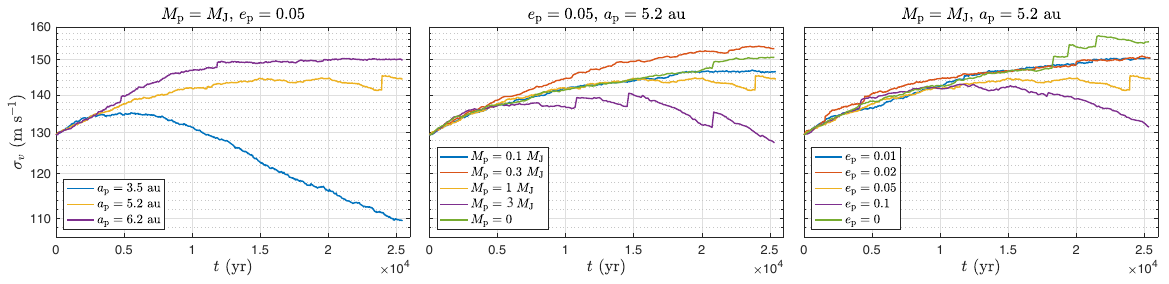}
    \caption{Similar to Fig. \ref{fig:m23_e_evo} but showing the time evolution of the random velocity (velocity dispersion) of planetesimals near 1 au.}
    \label{fig:m23_v_evo}
\end{figure*}

\begin{figure}
    \centering
    \includegraphics[width=0.48 \textwidth]{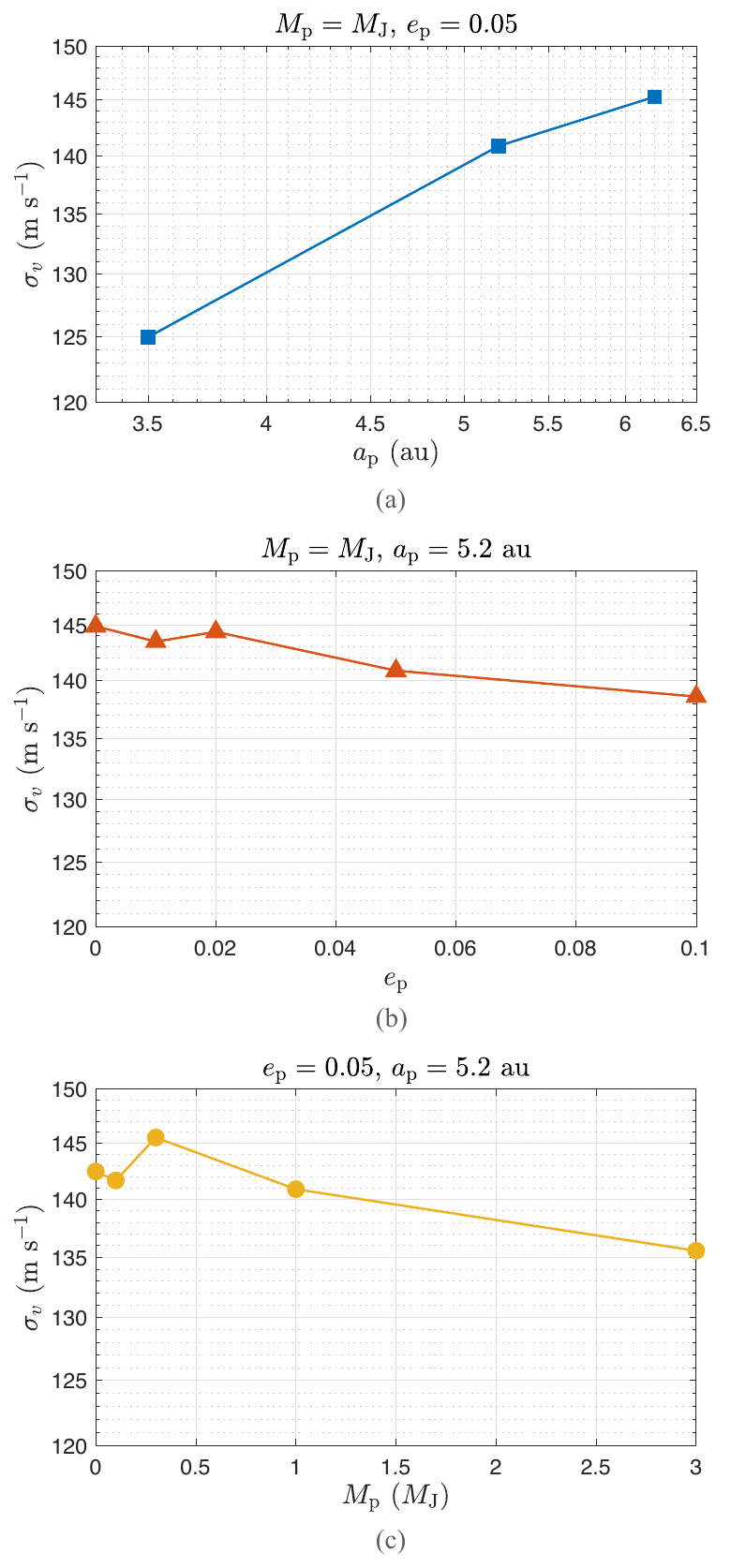}
    \caption{Dependence of the time-averaged random velocities of planetesimals near 1 au on the (a) location, (b) eccentricity, and (c) mass of the planet. The velocities are averaged over $\simeq 25000$ years.}
    \label{fig:m23_v_dependence}
\end{figure}

Finally, we show the time evolution of the random velocity of the planetesimals ($\sigma_v \simeq \sigma_e v_{\rm{K}}$) and its dependence on the mass, eccentricity, and location of the planet.
Fig. \ref{fig:m23_v_evo} shows the time evolution of the random velocities of the planetesimals located near 1 au in each model. 
The left, middle, and right panel each shows the dependence on the planet location, mass, and eccentricity respectively.
In Fig. \ref{fig:m23_v_dependence}, we take the time-averaged values of the random velocities in each model and show their dependence on the planet parameters more clearly. 
The velocity values are averaged over $\simeq 25000$ years.

Unlike in the small planetesimal case, even though a closely located giant planet (at 3.5 au) exerts strong perturbations on the inner planetesimals near 1 au, the accretion of these large planetesimals can actually be facilitated because their velocity dispersion decreases as a consequence of the combined effect of secular perturbation, gas damping, and dynamical friction.
When the mean eccentricities are raised due to secular perturbation, gas drag damps the dispersion of the eccentricity vectors and thus reduces the velocity dispersion.
In the mean time, dynamical friction between planetesimals prevents the differential precession of the eccentricity vectors, so that the velocity dispersion is not raised by the stronger secular perturbation. 
Such a trend is also observed when the planet mass or eccentricity is increased, although the decrease of $\sigma_v$ is almost negligible as these two parameters change.
When the planet is (a) located beyond $\simeq 5.2~$au, or (b) less massive than $\simeq M_{\rm{J}}$, or (c) less eccentric than $\simeq 0.05$, its impact on the random velocities of the planetesimals near 1 au is negligible.
In all the models including the mutual gravity of the planetesimals ("-L" and E0, M0), the velocity dispersion of planetesimals is always below the escape velocity ($v_{\rm{esc}} = 257.98~\rm{m}~s^{-1}$), indicating that the planetesimals can grow in a runaway fashion as in the non-perturbed environment.

These findings are good news for the accretion of terrestrial planets. 
If Jupiter has migrated inward during the early evolution of the Solar System as a result of planet-disk interaction, it would lower the random velocities of $\sim 100$-km-sized planetesimals near 1 au and assist the growth of the terrestrial planets.
A similar argument can be made for exoplanetary systems with similar configurations.

\section{Discussion} \label{sec:discussion}

\subsection{Terrestrial planet}

The results of our fiducial model is qualitatively consistent with those of KW00. 
However, because of the different model settings, our results of planetesimal relative velocities slightly deviate from those of KW00 within the same order of magnitude. 
In KW00, the inner disk is perturbed by both Jupiter and Saturn, while in our fiducial model, only Jupiter is in presence. 
Despite the smaller number of perturbing planets in our fiducial model, the slightly larger eccentricity of the planet ($e_{\rm{p}} = 0.05$ rather than 0.048, the present value of Jupiter as adopted in KW00) and the statistical method of calculating the relative velocities employed in our study are most likely to have caused the higher values of relative velocities in our results compared with those in KW00. 

According to KW00 and to our results for the fiducial model, planetesimal accretion is generally possible near 1 au, but the relative velocities can still be quite high ($\sim 100~\rm{m}~s^{-1}$) between populations of the largest mass ratio. 
Such high relative velocities can make accretion challenging for these bodies.
KW00 pointed out that the perturbations from Jupiter and Saturn would become considerably smaller if they were located further out at the time of the runaway growth of the inner cores, and then drifted inward to their present positions. 
Their results show that if Jupiter was at $a_{\rm{J}} = 6.2~$au and Saturn was at $a_{\rm{S}} = 10.5~$au, the relative velocities of planetesimals near 1 au would be significantly lower so that the growth of planetary embryos would not be handicapped by the perturbations from the two giant planets. 

Our results show that the perturbation can also be reduced if Jupiter was less massive or less eccentric than the present values during the runaway growth of planetesimals in the inner disk. 
The formation history of the Earth has been poorly constrained in previous studies so far.
According to \citet{Kleine2009}, about 63\% of Earth mass was accreted during  $\sim 10$-15 Myr after CAI formation. 
More recent studies have shown that our Solar System may have formed from rings of planetesimals created by pressure bumps rather than from a continuous disk \citep[e.g.][]{Izidoro_et_al_2021}.
Their model shows that in the inner ring ($\sim 1~$au) where the total mass in planetesimals is $\sim 2.5~M_{\oplus}$, planetesimals grow to roughly Mars-mass planetary bodies in $\sim 1$-3 Myr via planetesimal accretion.

The growth history of Jupiter is better constrained compared with that of the Earth. 
\citet{Kruijer_et_al_2020} has shown that Jupiter's core of 10-20$~M_{\oplus}$ accreted within $<0.5~$ Myr, while Jupiter 
reached $\sim 50~M_{\oplus}$ after $\sim 2~$Myr, and its final size of $\sim 318~M_{\oplus}$ before $\sim 4$-5 Myr.
Therefore, it is very likely that during the accretion of the planetesimals in the inner disk, the young Jupiter was much smaller and thus exerted much weaker perturbation on the planetesimals.
Apart from the mass, the eccentricity of Jupiter can also deviate from the present value. 
The eccentricity of Jupiter oscillates periodically as a result of secular interaction with other planets in the Solar System \citep[e.g.][]{Brouwer_vanWoerkom_1950, solar_system_dynamics}. 
If the young Jupiter had smaller eccentricity than the present value, it would also have weaker perturbation on  the inner planetesimals and thus lead to lower relative velocities.

The situation can be more optimistic if the initial sizes of planetesimals are larger (of a few hundred kilometers) as predicted by the streaming instability mechanism.
Our results show that for small planetesimals ($m \lesssim 10^{19}~$g) whose escape velocity and growth limits are low, the presence of a CJ would very likely prevent the collisional growth of objects with different sizes.
When a Jupiter-like planet is in presence, runaway growth of planetesimals is only allowed for equal-size bodies with masses larger than $\sim 10^{18}~$g. 
However, when larger planetesimals ($m = 10^{23}~$g) are considered, we find that the random velocities of the planetesimals are always smaller than the escape velocity (approximately by a factor of 2) in our models, enabling the runaway growth of these massive planetesimals under the perturbation of an external giant planet.

In the meantime, the inward migration of Jupiter (to an innermost orbit of 3.5 au) would further align the orbits of the planetesimals near 1 au and reduce their random velocities, enhancing the accretion of the embryos of terrestrial planets.

\subsection{Implications on exoplanetary systems}
By varying the mass, the eccentricity, and the location of the outer planet, we generalize the results of KW00 to extrasolar planetary systems.
Statistical estimates have shown that Jupiter-like planets are rare.
\citet{Raymond_2018} reported that only 1\% of Sun-like stars host Jupiter-like planets with orbital radii larger than 2 au and eccentricities smaller than 0.1. 
On the other hand, recent observations reported that the existence of cold giant planets and inner super-Earths is correlated to some degree.
It has been found that systems with super-Earths are more likely to harbor cold giant planets ($>1~$au)\citep{Bryan_2019, Zhu_2018}.
A quantitative estimate of such a correlation has been provided in \citet{Rosenthal_2022} as 40\% of cases. 
Such a correlation implies that the presence of an outer giant planet promotes (or at least does not inhibit) the formation of inner planets. 
This is supported by our results showing that the presence of a Jupiter-like planet does not affect the collisional growth of equal-mass, small planetesimals.
A smaller cold giant planet would be more lenient to planetesimal accretion, allowing runaway growth of objects with a higher mass ratio.
When the initial mass of planetesimals is large (corresponding to a few hundred km in size), the inward migration of the giant planet can even reduce the random velocity of the planetesimals by inducing more efficient gas damping on their orbits. 
Therefore, our results support the positive correlation between super-Earths and cold giant planets.

\subsection{Relative velocities}

The implications on planetary systems from our results are based on the comparison between the planetesimal relative velocities and the growth limits given in previous studies.
The uncertainty of these two quantities can lead to different planet parameters as growth conditions.

As we discussed in GK22, the growth limits derived using the critical impact energy required for catastrophic disruption $Q^*$ depend on multiple factors which can be oversimplified in our models. 
In addition, the relative velocities in our study are calculated statistically using the dispersion of the eccentricity vectors of planetesimals instead of recording the actual encounter velocity between two colliding planetesimals. 
Such a calculation method might lead to slightly higher results compared to those of KW00.
Comparing our results of the fiducial model shown in Fig. \ref{fig:v_compare_1au} and those in Fig. 6 in KW00, we find that our calculation method results in slightly higher values of relative velocities, even though we do not include the perturbation from Saturn. 
However, such discrepancies caused by the calculation method of relative velocities do not have a qualitative impact on our results. 

\subsection{Size distribution of planetesimals}

In section \ref{subsec:new_results}, we present the results of models considering large, equal-size planetesimals.
This is because when the mutual gravity of planetesimals is taken into account, we cannot approximately calculate the relative velocity between different-size bodies using the simple relation $\Delta v \simeq \Delta \boldsymbol{e} v_{\rm{K}}$ as in models where we consider small planetesimals and neglect their mutual gravitational interactions ("-S").
In order to reduce the computational cost and reduce the parameter space, we simply use equal-mass planetesimals to demonstrate the effect of varying the location of the planet in a relatively ``realistic'' scenario where the planetesimals interact with each other. 
However, in reality, it is unlikely that the planetesimals have a uniform mass.
Since the effect of the gravitational interaction of planetesimals is not the key point to be shown in this paper and will be further investigated in future work, we simply infer the possible influence of introducing a size distribution of planetesimals in our current models from previous research and our results.

Previous studies and our results from models of small planetesimals ("-S") have shown that the size-dependent orbital phasing due to the coupled effect of gas damping and secular perturbation would raise the relative velocities between objects with different sizes. 
However, this is a conclusion drawn from the analysis of non-interacting bodies. 
\citet{IDA199228} studied the velocity distribution of self-interacting planetesimals with a 4-component power-law mass distribution and found that the root mean square of the eccentricity and inclination of large planetesimals is reduced by the dynamical friction from small planetesimals; for a protoplanet embedded in a swarm of small planetesimals, the large eccentricity and inclination of the protoplanet immediately decrease due to the dynamical friction from the planetesimals and fluctuate around an equilibrium state.
From these conclusions, we can infer that, if a size/mass distribution is introduced in our models ("-L" and E0, M0), given that the total mass of the small bodies is much larger than that of the large bodies, the eccentricity vectors of large planetesimals would converge towards those of the smaller planetesimals as a result of dynamical friction.
In this way, the size-dependent orbital phasing would be mitigated by the gravitational viscosity of the self-interacting planetesimal swarm.
This corollary will be tested with numerical experiments in our future paper.

\section{Summary} \label{sec:summary}

\subsection{Conclusions}
We explore the dynamics of planetesimals in the inner disk under the coupled effect of gas drag and perturbation from a giant planet in the outer disk. 
We find qualitatively consistent results with previous studies: under the perturbation from the planet, gas drag induces orbital alignment for planetesimals of equal size and thus reduces the relative velocities of planetesimals of similar sizes; for planetesimals of high mass ratio, the presence of gas drag prevents the alignment of orbits and keeps their relative velocities high. 
When the perturbing planet resembles Jupiter, collisional growth of similar-size planetesimals is allowed in the terrestrial planet forming region (near 1 au), while the high relative velocities of planetesimals in the asteroidal region inhibit planetesimal accretion and growth of  planet embryos there.

By varying the mass, eccentricity, and location of the planet, we generalize our results to extrasolar planetary systems. 
Generally speaking, a larger mass and eccentricity or a smaller orbital distance of the planet leads to higher planetesimal relative velocities and less efficient accretion of planetary embryos.
For a planet with $e_{\rm{p}} = 0.05$, a mass exceeding Jupiter mass would inhibit the runaway growth of planetesimals near 1 au and thus prevent the formation of terrestrial planets; a small mass below $\simeq 0.1~M_{\rm{J}}$ would allow planet formation even near 2.6 au, which corresponds to the asteroidal region in the Solar System. 
A Jupiter-mass planet with an eccentricity of $\gtrsim 0.05$ would likely impede the growth of terrestrial planets near 1 au; when it is on a nearly circular orbit with $e_{\rm{p}} \lesssim 0.01$, the growth of planet embryos is generally safe within 3 au, except at the 2:1 resonance at $a\simeq 2.5~$au. 
When a Jupiter-like planet is located at $\simeq 3.5$ au, basically only equal-size bodies can undergo collisional growth; when the planet is located further out, collisional growth is allowed for more values of target-projectile mass ratio.

When considering larger planetesimals of a few hundred km in size ($10^{23}~$g), we find that runaway growth near 1 au is always allowed when a Jupiter-like planet is located beyond 3.5 au, which is the minimum semi-major axis of the planet considered in our model. 
Within the parameter range in our study, when the planet is located on a closer orbit, the random velocities of the planetesimals are reduced compared to the case when the planet is located further out, because the mean eccentricity is raised to a higher value due to larger forced eccentricity.
As a result, the gas-damping effect becomes more efficient, which leads to smaller velocity dispersion.
These results indicate that the inward migration of the giant planet could facilitate the growth of inner rocky planets, provided that it starts from accretion of initially large planetesimals ($m\simeq 10^{23}~$g).
Our results imply that if planetesimals form via streaming instabilities and if the gas disk is massive enough, the presence of a cold giant planet can assist the formation of inner rocky planets by reducing the relative velocities of planetesimals during its inward migration. 
Such a mechanism provides a possible explanation for the correlation between inner super-Earths and cold giant planets.

\subsection{Limitations and future perspectives}
A major limitation of this work stems from the simplified model setup. 
The gas component of the disk is assumed to be axisymmetric and circular. 
However, as we have discussed in Section 4.4 in GK22 and in Section \ref{subsec:gas_disk}, the gas disk can become eccentric under the perturbation of a massive planet or a stellar companion.
An eccentric gas disk can raise the relative velocity of planetesimals and makes the environment more hostile for planet formation (see Section \ref{subsec:gas_disk}.

Apart from the gas component, the initial distribution of the solid component of the disk in the small planetesimal models is also simplified - we only employ a power-law distribution of the planetesimals without considering any structures in the planetesimal disk, such as rings or gaps, which results in discontinuous distribution of the solid surface density.
Recent studies have shown that planets could have formed from rings of planetesimals rather than a continuous disk \citep[e.g.][]{Izidoro_et_al_2021}.
By considering the concentration of planetesimals in rings instead of a continuous disk, we might expect faster growth of planetary embryos at the ring locations owing to the higher solid surface density.
Similarly, we consider a simplified size distribution of planetesimals by assigning the same number of particles to each particle mass, and the particles do not interact with each other (no inter-particle gravity).
In models of large planetesimals ("-L" and E0, M0), we consider planetesimals initially distributed in a ring and include the mutual interaction.
However, due to the high computational cost, we currently limit the number of particles to 2048, which is insufficient to produce an Earth-like planet.
In addition, the width of the planetesimal ring $\Delta a$ also affects the relative velocities in our calculation. 
In future models, we will increase the number of particles to follow the more realistic dynamical evolution of planetesimals, and the effect of the mutual gravitational interaction of planetesimals on orbital alignment will be investigated in more detail.

Finally, our current models do not simulate the migration of the planet. 
In our future models, we will implement a migration prescription for the planet to account for a more realistic formation scenario of the outer giant planet as well as the redistribution of solid materials in the disk (e.g., scattering and shepherding of planetesimals by the planet).

Owing to the uncertainties in our understanding of initial conditions  in a protoplanetary disk, we recognize our investigation using simplified models as a first step towards a more sophisticated picture of planet formation.

\vspace{12pt}
\noindent \textit{Acknowledgements}
\vspace{6pt}

We thank the anonymous reviewer for his/her constructive comments and suggestions helping us to improve this work.
The numerical computations of models were conducted on the general-purpose PC clusters and GPU clusters at the Center for Computational Astrophysics, National Astronomical Observatory of Japan. 
E.K. is supported by JSPS KAKENHI grant No. 18H05438.

\bibliography{references}{}
\bibliographystyle{aasjournal}

\end{document}